\documentclass[12pt,preprint]{aastex}

\begin{document}

\title{\large \bf Repeated injections of energy in the first 600 ms \\
of the giant flare of SGR 1806-20}

\author{Toshio Terasawa$^{1}$, Yasuyuki T. Tanaka$^{1}$, Yasuhiro Takei$^{1}$, 
Nobuyuki Kawai$^{2}$, Atsumasa Yoshida$^{3}$, Ken'ichi Nomoto$^{4}$, Ichiro Yoshikawa$^{1,5}$, 
Yoshifumi Saito$^{5}$, Yasumasa Kasaba$^{5}$, Takeshi Takashima$^{5}$, Toshifumi Mukai$^{5}$, 
Hirotomo Noda$^{6}$, Toshio Murakami$^{7}$, Kyoko Watanabe$^{8}$, Yasushi Muraki$^{8}$, Takaaki Yokoyama$^{1}$
$\&$ Masahiro Hoshino$^{1}$}

\affil{$^{1}$Department of Earth and Planetary Science, University of Tokyo, Tokyo 113-0033, Japan\\
$^{2}$Department of Physics, Tokyo Institute of Technology, Tokyo, 152-8551, Japan\\
$^{3}$Department of Physics and Mathematics, Aoyama Gakuin University,
Sagamihara, 229-8558, Japan\\
$^{4}$Department of Astronomy, University of Tokyo, Tokyo 113-0033, Japan\\
$^{5}$ISAS/JAXA, Sagamihara, 229-8510, Japan\\
$^{6}$Mizusawa Astrogeodynamics Observatory, National Astronomical Observatory,
Mizusawa, 023-0861, Japan\\
$^{7}$Department of Physics, Kanazawa University, Kanazawa, 920-1192, Japan\\
$^{8}$Solar Terrestrial Environmental Research Laboratory, Nagoya University,
Nagoya, Aichi 464-8601, Japan\\
to be published in {\it Nature} on 28 April 2005.}

\begin{abstract}
The massive flare of 27 December 2004 from the soft $\gamma$-ray repeater SGR 1806-20, a possible 
magnetar$^{1-3}$, 
 saturated almost all 
$\gamma$-ray detectors$^{4-7}$, 
meaning that the profile of the pulse was poorly 
characterized. An accurate profile is essential to determine physically what was happening at the source. 
Here we report the unsaturated $\gamma$-ray profile for the first 600 ms of the flare, with a time resolution 
of 5.48 ms. The peak of the profile (of the order of 10$^7$ photons cm$^{-2}$ s$^{-1}$) was reached $\sim50$ ms
 after the onset of the flare, and was then followed by a gradual decrease with superposed oscillatory 
modulations possibly representing repeated energy injections with $\sim60$ ms intervals. The implied total 
energy is comparable to the stored magnetic energy in a magnetar ($\sim10^{47}$ erg) based on the dipole 
magnetic field intensity ($\sim10^{15}$ G), suggesting either that the energy release mechanism was extremely 
efficient or that the interior magnetic field is much stronger than the external dipole field$^{2}$. 
\end{abstract}

~

At the onset of the giant flare of SGR 1806-20, plasma particle detectors on the Geotail 
spacecraft detected an extremely strong signal of soft $\gamma$-ray photon fluxes 
(integrated above $\sim50$ keV) during the initial intense phase of the giant flare 
($t$=0-600 ms). Figure 1 shows the count profiles of two detectors, $N_\mathrm{MCP}$ 
(red symbols) from the microchannel plates (MCPs), and $N_\mathrm{CEM}$ (blue symbols) 
from channel electron multipliers (CEMs), where $N_\mathrm{CEM}$ is scaled by a factor 
of 280 to account for the sensitivity difference. The onset time ($t$=0) corresponded 
to 21 h 30 min 26.35 s Universal Time (UT), which was consistent with the expected 
arrival time of the onset signal at the Geotail position. To understand how the flare 
energy release occurred, the detailed time profile of the flare is important. Before the 
onset $N_\mathrm{MCP}$ was at the background level ($<\sim$11 counts, shown by a black arrow), and then 
increased to 839 counts within 5.48 ms, so that the e-fold time of the initial rise was 
shorter than $\sim1.3$ ms. After the intermediate level of 1,330 counts at $t$=5.48 ms, 
the MCPs were saturated and $N_\mathrm{MCP}$ could not be determined until $\sim176$ ms. 
Between $t$=22.7 and 170 ms, we could obtain $N_\mathrm{CEM}$ values instead. The scaled 
$N_\mathrm{CEM}$ increased to 25,900 at $t$=33.6 ms, thus giving an e-fold time of 
9.5 ms. (A data gap between $t$=11.0 and 22.7 ms shown 
by the leftmost grey bar was due to 
the scheduled instrumental operation and not caused by the flare itself.) Between $t$=33.6 and 55.5 ms the scaled 
$N_\mathrm{CEM}$ stayed at the peak level of $\sim25,000$-$27,000$. After $t$=61 until 170 ms the scaled $N_\mathrm{CEM}$ 
decreased gradually with oscillatory modulation, which suggests repeated energy injections at $\sim60$ ms intervals.
(Note that a similar injection profile was also seen during the
impulsive phase of the giant flare of SGR0526-66 on 5 March 1979 (ref.8).)
After 
176 ms $N_\mathrm{MCP}$ became available again and showed a continuing exponential decay with an e-fold time of $\sim66$ 
ms until $t$=380 ms. The decay profile of scaled $N_\mathrm{CEM}$ available for $t$=210-308 ms is consistent with that of 
$N_\mathrm{MCP}$. Between $t$=397 and 500 ms several humps were seen on the profile of $N_\mathrm{MCP}$. Although 
less significant, scaled $N_\mathrm{CEM}$ showed a similar hump for $t$=402-451 ms. (Note that the same humps were 
also detected by the BAT detector on the Swift spacecraft$^{7}$.) 
The physical origin of these humps is not clear 
at the moment, but may represent some additional energy-releasing process. After $t$=470 ms, $N_\mathrm{MCP}$ again 
decayed with an e-fold time of $\sim57$ ms.

To convert the observed count rates to physical quantities such as energy flux, we need the energy spectrum information, 
which was not available from the Geotail observation alone. We have therefore taken three reported function forms at the 
peak of the giant flare$^{4,5,7}$ 
(Table 1) and integrated them above 50 keV. We then found that the resultant estimations 
of photon number flux, energy flux and fluence (for $t$=0-600 ms) are almost independent of the choice of the energy 
spectrum, and are $\sim2.5\times10^7$ photons cm$^{-2}$ s$^{-1}$, $\sim20$ erg s$^{-1}$ cm$^{-2}$ and $\sim2$ erg 
cm$^{-2}$, respectively. The corresponding total energy radiated from SGR 1806-20 is estimated to be $\sim5\times10^{46}$ 
($\Omega$/4$\pi$) d$_{15}^2$ erg, where d$_{15}$ is the distance scaled by 15 kpc and $\Omega$ is the solid angle of the 
radiation. Here we note that the solid-angle factor ($\Omega$/4$\pi$) is not likely to be as small as 10$^{-2}$, as is typically 
assumed for relativistic jets for GRBs. ($\Omega$/4$\pi$)$>0.1$ is more likely for the intense initial spikes of SGRs because 
they have been seen in all the three giant flares of SGRs that could have been detected without them. Therefore the 
presence of a very efficient mechanism is implied, which promptly releases (on a timescale of $\sim60$ ms) a major fraction 
of  the stored magnetic energy in a magnetar, $E_\mathrm{mag}\approx1.7\times10^{47}$B$_{15}^2$R$_6^3$ erg
 (where B$_{15}$ and R$_6$ are the internal magnetic field scaled by 10$^{15}$ G 
and the radius of the magnetar scaled by 10$^6$ cm). Alternatively, as suggested by 
ref. 2, the internal magnetic field could be as strong as (5-10)$\times10^{15}$ G so as to 
permit the emission of multiple giant flares over the lifetime of a magnetar.

As we noted above, there were humps in the light curve of the 2004 giant flare at 
$t$=400-500 ms. Similar humps were also observed$^{9}$ 
at 200-600 ms after the onset of 
another giant flare of SGR 1900+14 on 27 August 1998, whose total energy is smaller by 
a factor of $\sim100$ than the 2004 giant flare. From similarities in the timings of 
the humps despite the large difference in the total energies, we suggest that the humps 
more probably represented continuing energy injections, rather than the results of 
interactions of the flare ejecta with environmental matter.

\section*{Methods}

The Low Energy Particle (LEP) experiment$^{10}$ 
aboard Geotail consists of an ion detector 
with seven independent MCPs and electronics systems, and an electron detector with seven 
independent CEMs and electronics systems, both of which are designed to measure plasma 
particles in the solar wind and magnetospheric environment. When the giant flare 
occurred, Geotail was at (-1.5997$\times10^5$,-97945,-19671) km using the 
Geocentric Solar Inertia (GCI) coordinates (J2000), which was in the solar wind about $\sim10$ 
earth radii upstream from the bow shock. Although MCPs and CEMs kept measuring the solar 
wind ions and electrons throughout the giant flare interval, these particles were being 
selected electrostatically and came into the detectors mainly at some limited timings 
that fortunately did not overlap the giant flare interval: the contribution of solar wind 
ions to $N_\mathrm{MCP}$, which is the sum of counts over seven MCPs, was at most 30, 
and thus was negligible for the study of intense $\gamma$-ray photons. On the other hand, 
during subintervals ($t<10$ ms, 175 ms$<t<$200 ms, 320 ms$<t<$400 ms and 450 ms$<t<$600 ms)
, $N_\mathrm{CEM}$, which is the sum of counts over seven CEMs, was affected by solar 
wind electrons, and was not available for the $\gamma$-ray photon detection.

Another fortunate factor was that the angular distance between the Sun and SGR 1806-20 
was $\sim5$ degrees, so that previous knowledge of the `calibration' of the LEP detector 
to be used as a soft $\gamma$-ray photon counter on the basis of solar flare photon 
analysis was directly applicable to the interpretation of the observed characteristics 
of photons from SGR 1806-20. By comparing the count rates of MCP and CEM with the hard 
X-ray$^{11}$ and $\gamma$-ray$^{12}$ data from the Yohkoh space solar observatory during 
major solar flares in 1997-2001 (ref.13), we have seen that MCP and CEM are sensitive to 
soft $\gamma$-ray photons above $\sim50$ keV, where their sensitivities are evaluated as 
the product \textit{[$\epsilon\,$S]} of quantum efficiency $\epsilon$ and effective 
detection area $S$ summed over seven MCPs and seven CEMs against $\gamma$-ray photons. 
(Here $\epsilon$ is defined to include not only the detector response itself but also 
the attenuation factor inside the spacecraft.) From the spectral information provided by 
the Yohkoh observations, we have calculated photon fluxes in two energy ranges, 
L (50-500 keV) and H (above 500 keV), and then estimated 
\textit{[$\epsilon\,S$]}$_\mathrm{MCP,L}$ and \textit{[$\epsilon\,S$]}$_\mathrm{MCP,H}$ 
separately for these two energy ranges. (This separation is possible because the energy 
spectra of incident solar $\gamma$-ray photons differ from event to event.) Figure 2 
shows the calibration result. Along the line of sight towards SGR 1806-20, we have found 
that \textit{[$\epsilon\,S$]}$_\mathrm{MCP,L}$=(0.19$\pm$0.06) cm$^2$ and 
\textit{[$\epsilon\,S$]}$_\mathrm{MCP,H}$=(0.22$\pm$0.16) cm$^2$ where systematic errors 
are included. The estimation of \textit{[$\epsilon\,S$]}$_\mathrm{CEM}$ has been also 
done with solar flare $\gamma$-ray photons, and the result is summarized as 
\textit{[$\epsilon\,S$]}$_\mathrm{CEM}$$\approx$1/280 of 
\textit{[$\epsilon\,S$]}$_\mathrm{MCP}$. The smallness of \textit{[$\epsilon\,S$]}$_\mathrm{CEM}$ 
as compared with \textit{[$\epsilon\,S$]}$_\mathrm{MCP}$ is consistent with the 
difference in physical sizes of CEMs (millimetres) and MCPs (several centimetres).

It is noted that \textit{[$\epsilon\,S$]}$_\mathrm{MCP}$ obtained above is by a factor 
$10^{-2}$-$10^{-3}$ smaller than those of conventional $\gamma$-ray detectors. 
Nonetheless MCPs were saturated during $\sim150$ ms of the onset of the giant flare 
of SGR 1806-20. We have made that the dead-time analysis for MCPs and found that the 
characteristic dead time is $\sim4.3$ $\mu$s, which is consistent with the pre-flight 
calibration of the LEP system as well as the calculated circuit time constant. 
$N_\mathrm{MCP}$, shown in Fig. 1, is after the dead-time correction, which becomes 
significant above $\sim1,000$ counts. On the other hand, CEMs, 
which are two orders of magnitude less sensitive than MCPs, were found to be free from 
the saturation effect even at the peak of the giant flare.

\section*{Acknowledgments}

We thank R. Yamazaki, K. Ioka, K. Shibata, T. Minoshima and K. Hurley for comments and
discussions. This work was supported in part by grants
from the Ministry of Education, Culture, Sports, Science and Technology of Japan.

\par\noindent
Correspondence and requests for materials should be addressed to T. T. (e-mail:
terasawa@eps.s.u-tokyo.ac.jp).

\newpage
\begin{table}[htbp]
\begin{center}
\begin{tiny}
\begin{tabular}{cccccc} \\ \hline
Model & Photon flux  & Energy flux  & Fluence  & Isotropic luminosity  & Isotropic total energy \\ [-4pt]
      & (photons cm$^{-2}$ s$^{-1}$) &  (erg cm$^{-2}$ s$^{-1}$) &  (erg cm$^{-2}$) &  (erg s$^{-1}$) &   (erg)  \\ \hline
Planck & $(2.5^{+1.1}_{-0.6})\times107$ & $19^{+9}_{-4}$ & $2.0^{+0.9}_{-0.5}$ & $(5.1^{+2.3}_{-1.2})\times10^{47}d_{15}^2$ & $(5.4^{+2.4}_{-1.3})
\times10^{46}d_{15}^2$ \\
$E^{-0.2}\mathrm{exp}(-E/480)$ & $(2.5^{+0.9}_{-0.5})\times107$ & $18^{+7}_{-4}$ & $1.9^{+0.7}_{-0.45}$ & $(4.9^{+1.7}_{-1.0})\times10^{47}d_{15}^2$ 
& $(5.2^{+1.8}_{-1.1})\times10^{46}d_{15}^2$ \\
$E^{-0.7}\mathrm{exp}(-E/800)$ & $(2.5^{+0.8}_{-0.5})\times107$ & $18^{+6}_{-3}$ & $1.9^{+0.6}_{-0.4}$ & $(4.9^{+1.4}_{-1.0})\times10^{47}d_{15}^2$ & 
$(5.1^{+1.6}_{-0.9})\times10^{46}d_{15}^2$ \\ \hline
\end{tabular}
\end{tiny}
\caption{Estimations of energy flux, fluence, luminosity and total energy for several models. 
(The Planck distribution with $kT$=175 keV is from ref. 4; 
the power-law distributions with 
exponential cut-offs at 480 keV and 800 keV are from refs. 7 and 5, 
respectively.)} 
\end{center}
\end{table}

\newpage
\begin{figure}
\plotone{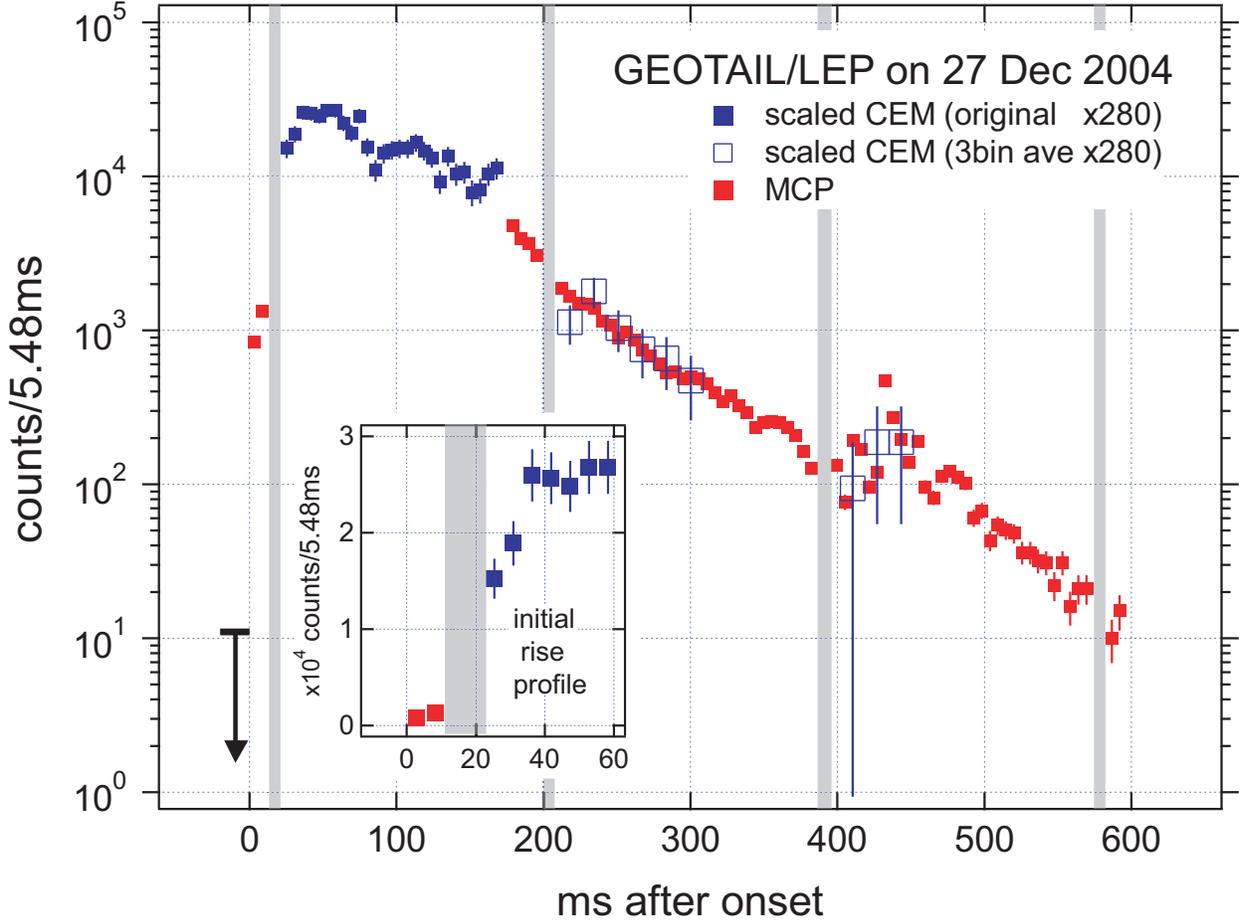}
\caption{\label{fig:lc}
Observed photon counts during the first 600 ms of the giant flare.
$N_\mathrm{MCP}$ (red squares) and $N_\mathrm{CEM}$ (blue solid squares) show the counts of MCP
and CEM instruments accumulated over a bin of 5.48 ms duration. Averages
over three successive bins (16.44 ms) are taken for $N_\mathrm{CEM}$ (blue open squares) after
200 ms. Vertical error bars represent 1 $\sigma$ statistical deviations, which become
smaller than the symbol size for $N_\mathrm{MCP}$ $>100$. Grey bars show the data gaps
every 32 bin (or every 187.09 ms) caused by the scheduled instrumental
operation.
Before $t$=0, the photon counts were below $\sim$11. The black arrow shows this upper limit. 
The inset shows $t$=0-60 ms with a linear scale.
}
\end{figure}

\newpage
\begin{figure}
\plotone{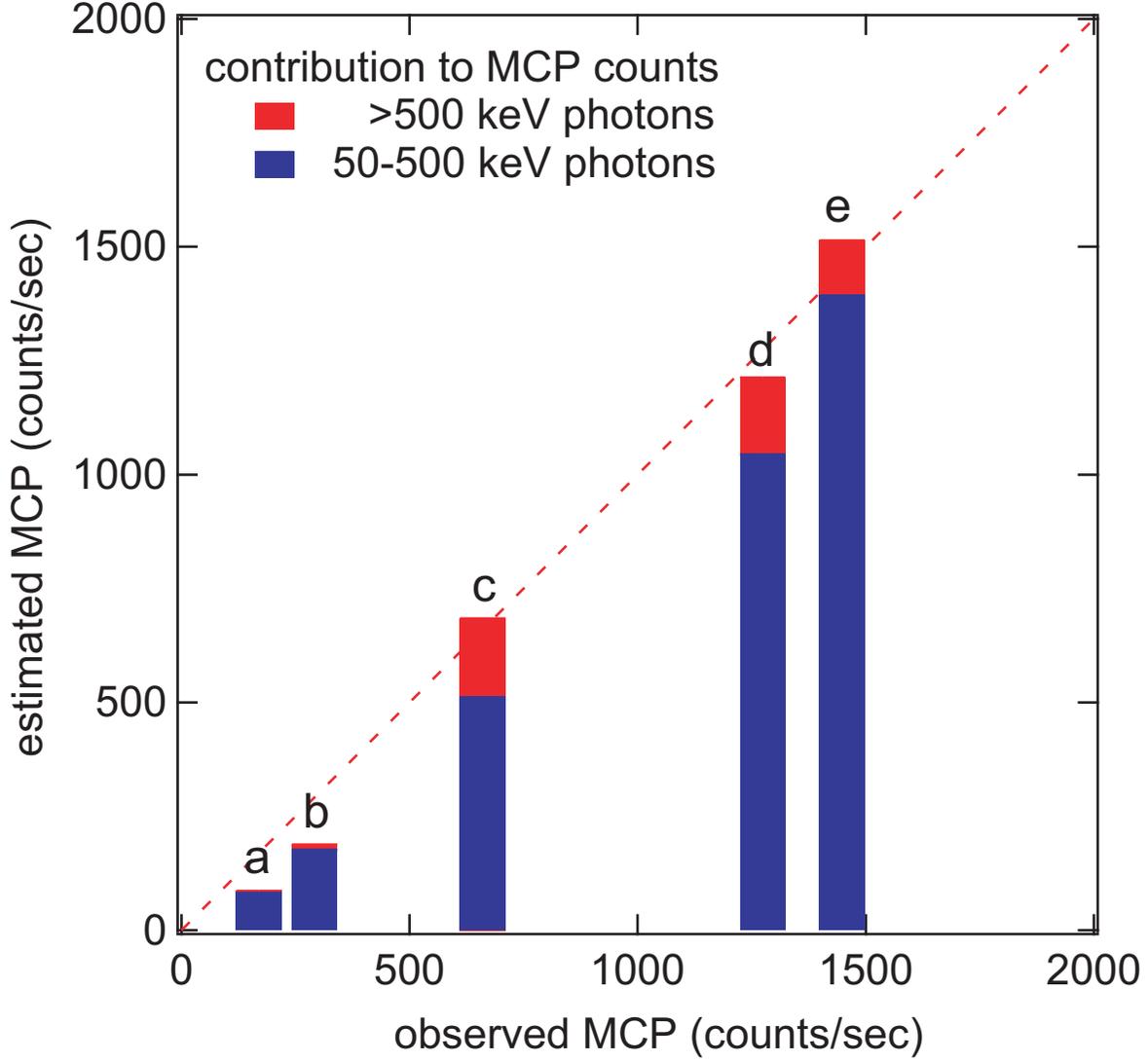}
\caption{\label{fig:calib}
Calibration of the $\gamma$-ray sensitivity of the MCP detector. The
observed MCP counts (counts per s) are compared with the counts
synthesized from the solar $\gamma$-ray photon observations with the estimated
\textit{[$\epsilon\,S$]}$_\mathrm{MCP,L}$ and \textit{[$\epsilon\,S$]}$_\mathrm{MCP,H}$ 
for five solar $\gamma$-ray flares (Goes class X3.7 on
22 November 1998, X4.9 on 18 August 1998, X2.3 on 24 November 2000, X5.3 on 25
August 2001 and X9.4 on 6 November 1997). The blue and red parts of each bar
represent contributions from the photons in the energy ranges of 50-500 keV
and above 500 keV, respectively.
}
\end{figure}

\end{document}